# Course Difficulty Estimation Based on Mapping of Bloom's Taxonomy and ABET Criteria

Premalatha M[1][0000-0001-7685-5417] Suganya G[2][0000-0001-9560-4760] Viswanathan V[3][0000-0001-9996-0308] and

G Jignesh Chowdary[4][0000-0002-3738-3960]

[1,2,3,4] Vellore Institute of Technology, Chennai 600127, India
suganya.g@vit.ac.in

**Abstract.** Current Educational system uses grades or marks to assess the performance of the student. The marks or grades a students' scores depends on different parameters, the main parameter being the difficulty level of a course. Computation of this difficulty level may serve as a support for both the students and teachers to fix the level of training needed for successful completion of course. In this paper, we proposed a methodology that estimates the difficulty level of a course by mapping the Bloom's Taxonomy action words along with Accreditation Board for Engineering and Technology (ABET) criteria and learning outcomes. The estimated difficulty level is validated based on the history of grades secured by the students.

**Keywords:** Course Difficulty, Bloom's action words, ABET Criteria, Student Learning Outcomes.

## 1 Introduction

### 1.1 Course Difficulty

Course difficulty estimation [1][2] has been a problem for many ages. A single parameter might not be sufficient in estimating the difficulty of a course. Course difficulty is based on various factors. It depends on the student's learning capacity, instructors' capability of delivering the lecture contents, course contents need more cognitive level thinking, courses might need higher-order thinking capability, etc. So obviously it affects the grades of the students. Less difficulty might yield great grades and more difficulty might yield lower grades of a student [2][3].

Many technics and methodology such as bloom's taxonomy [4] have been defined to help to assist us in predicting course yet it has not been precise due to having human errors in it as all technics require human work. Humans cannot provide enough confidence level in the predictions and require more sophisticated mathematical models for prediction. Prediction of course difficulty is vital to understand the quality of a particular course and to study the development of each generation of students. These data can further help us to improve a course based on the development seen in a different generation of students

<div align="center">

1.2    **Prerequisite Idea**

</div>

### 1.2.1    Bloom's Taxonomy

Bloom's taxonomy [4][5] is a set of three hierarchical models used to classify educational learning objectives into levels of complexity and specificity.

The three lists cover the learning objectives in cognitive, affective and sensory domains. The cognitive domain list has been the primary focus of most traditional education and is frequently used to structure curriculum learning objectives, assessments and activities.

There are six major categories in the cognitive domain:

i.    Remembering: Remembering involves recognizing or remembering facts, terms, basic concepts, or answers without necessarily understanding what they mean. Its characteristics may include:

ii.   Comprehension: Comprehension involves demonstrating an understanding of facts and ideas by organizing, comparing, translating, interpreting, giving descriptions, and stating the main ideas.

iii.  Application: Applying involves using acquired knowledge—solving problems in new situations by applying acquired knowledge, facts, techniques and rules. Learners should be able to use prior knowledge to solve problems, identify connections and relationships and how they apply in new situations.

iv.   Analysis: Analyzing involves examining and breaking information into component parts, determining how the parts relate to one another, identifying motives or causes, making inferences, and finding evidence to support generalizations.

v.    Synthesis: Synthesizing involves building a structure or pattern from diverse elements; it also refers to the act of putting parts together to form a whole

vi.   Evaluation: Evaluating involves presenting and defending opinions by making judgments about information, the validity of ideas, or quality of work based on a set of criteria

### 1.2.2    ABET Criteria

Accreditation Board for Engineering and Technology (ABET) recognizes and supports the prerogative of institutions to adopt and use the terminology of their choice, it is necessary for ABET volunteers and staff to have a consistent understanding of terminology.

The program must have documented student outcomes that prepare graduates to attain the program educational objectives. ABET Criterion 3[22] Student outcomes are outcomes (a) through (k) plus any additional outcomes that may be articulated by the program. Figure 1 depicts the student outcomes followed by B.Tech CSE programme (Curriculum 2014) at Vellore Institute of Technology, Chennai, India.

(a) an ability to apply knowledge of mathematics, science, and engineering
(b) an ability to design and conduct experiments, as well as to analyze and interpret data
(c) an ability to design a system, component, or process to meet desired needs within realistic constraints such as economic, environmental, social, political, ethical, health and safety, manufacturability, and sustainability
(d) an ability to function on multidisciplinary teams
(e) an ability to identify, formulate, and solve engineering problems
(f) an understanding of professional and ethical responsibility
(g) an ability to communicate effectively
(h) the broad education necessary to understand the impact of engineering solutions in a global, economic, environmental, and societal context
(i) a recognition of the need for, and an ability to engage in life-long learning
(j) a knowledge of contemporary issues
(k) an ability to use the techniques, skills, and modern engineering tools necessary for engineering practice.
(l) An ability to apply mathematical foundations, algorithmic principles and computer science theory in modeling and design of computer-based systems (CBC)
(m) An ability to apply design and development principles in the construction of software systems (CS)

Figure 1: ABET Criteria a-m

### 1.2.3    Mapping ABET Criteria students' outcomes with Blooms' Taxonomy

Blooms' Taxonomy is used for evaluating the higher order thinking and lower order thinking capacity of a student for answering a question during examination. Bloom's is not only for evaluating the question but it is for classroom teaching too[6]. Blooms' Taxonomy is mapped with ABET criteria so that a course can be defined with the level of cognitive thinking needed while learning the course.

In the proposed method, difficulty of all courses in the curriculum, by mapping the action words of bloom with the ABET criteria. The sequence of contents discussed in this paper is as follows: section 2 presents a detailed analysis on Literature, section 3 deal with Methodologies and Implementation, Results and Discussions are elaborated on section 4 and section 5 concludes the work.

## 2    Literature Survey

Difficulty of a course is estimated by considering the average grades awarded, rank correlation coefficient (rho) – means, scaling analysis and cluster analysis as factors. [2][12]. A study of item difficulty assessment says that it depends of the learners, course contents and the scores students have got with respect to Apriori algorithm, Subject Matter Expert [7]. Impact of course difficulty in students' performance is studied in [8], which analyses the previous student' results and predicted the results of the current student' using AdaBoost, J48 by AdaBoost, M5P. Based on the learning dependency, difficulty of knowledge units are ranked with respect to subjective difficulty and objective difficulty [9]. Some research discusses on what way rating of instructors [10] and feedback [11] by student helps the instructors to improve their performance in examinations.

Course planning [13][14][15] is a recommender system in which the students are recommended with sequence of courses to be taken or registered for each semester. Based on the interest of the students, some courses are recommended [16] in which Course recommendations are based on user profiles from users' interest description, browse log and subscriptions. Learning objects are also recommended sequentially [17] in which the students are recommended with the learning path and advised with how a course contents has to be sequentially learnt.

A personalized learning route is suggested to learn the sequence of learning object and if the student fails in the assessment during the learning process, the route will be modified/repaired and recommended with new objectives. Blooms' Taxonomy is used for evaluating the question paper's cognitive thinking [19] Students' course Outcomes are used for assessing the performance of a student in their examination based on the difficulty index of a course. With the help of bloom, final exam papers are evaluated and a difficulty index is identified [20]. A cognitive map [21] is provided to the research scholars which gives them a clear picture of how to start and to proceed with their research.

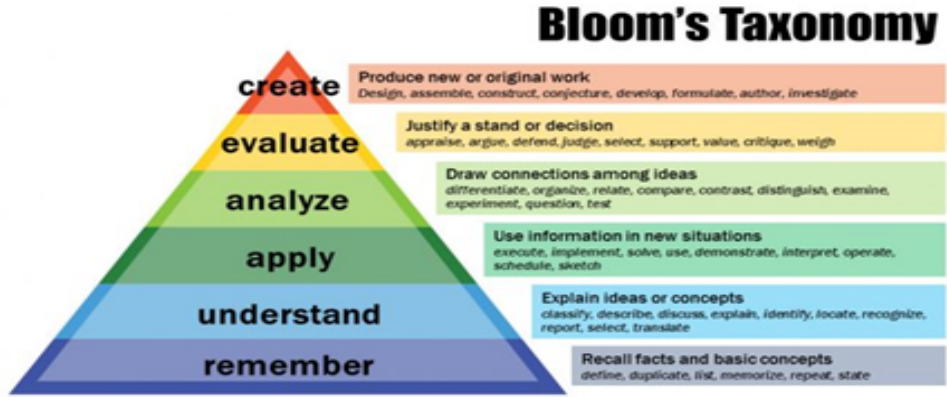

Figure 2: Bloom's Taxonomy

SUMMARY OF THE BLOOM'S TAXONOMY

| Level of Complexity | Cognitive (Knowledge) | Affective (Attitude) | Psychomotor (Skill) |
|---|---|---|---|
| 1 | Knowledge | Receiving | Imitation |
| 2 | Comprehension | Responding | Manipulation |
| 3 | Application | Valuing | Precision |
| 4 | Analysis | Organizing | Articulation |
| 5 | Synthesis | Characterizing by value or value concept | Naturalization |
| 6 | Evaluation | | |

Figure 3: Level of complexity of Bloom's Taxonomy [21]

## 3    Implementation

### 3.1    ABET criteria mapping with Blooms' Taxonomy

In this method, Bloom's Taxonomy is specified with hierarchy and identified with every ABET criteria of student's outcomes is mapped with the action words of Bloom's Taxonomy and arrived with a level of complexity [21]. Figure 2 depicts the bloom's taxonomy hierarchy and Figure 3 depicts the level of complexity applied for each category of Bloom's taxonomy.

By considering the Bloom's complexity and based on the action keywords of bloom's every ABET criteria is mapped with the level of complexity of bloom as specified in Table 1.

### 3.2    Estimation of difficulty for ABET criteria a-m

Every criteria of ABET is mapped with the level of complexity of bloom and arrived with the difficulty of the criteria as specified in table 1.Sample Heading (Third Level). Only two levels of headings should be numbered. Lower level headings remain unnumbered; they are formatted as run-in headings.

**Table 1: Applying level of complexity for ABET**

| ABET Criteria | Reme mber (1) | Unde rstand (2) | Appl y (3) | Anal yze (4) | Evalu ate (5) | Creat e (6) | Total (21) |
|---------------|---------------|-----------------|------------|--------------|---------------|--------------|------------|
| a | 1 | 2 | 3 |   |   |   | 6 |
| b | 1 | 2 | 3 | 4 | 5 | 6 | 21 |
| c | 1 | 2 | 3 | 4 | 5 | 6 | 21 |
| d | 1 | 2 | 3 |   |   |   | 6 |
| e | 1 | 2 | 3 | 4 | 5 | 6 | 21 |
| f | 1 | 2 |   |   |   |   | 3 |
| g | 1 | 2 |   |   |   |   | 3 |
| h | 1 | 2 | 3 |   |   |   | 6 |
| i | 1 | 2 | 3 | 4 | 5 | 6 | 21 |
| j | 1 |   |   |   |   |   | 1 |
| k | 1 | 2 | 3 |   |   |   | 6 |
| l | 1 | 2 | 3 | 4 | 5 | 6 | 21 |
| m | 1 | 2 | 3 | 4 | 5 | 6 | 21 |

### 3.3    Estimation of course difficulty

If a course is identified with a criterion, the difficulty of a course will be the sum of all difficulty index of the identified ABET criteria as specified in equation (1). If a course has a, h, k, l as criteria's, difficulty of that course is 6+6+6+21=39. They found difficulty index of a course is cross verified with the mean of the grades scored by the past three generations of students.

$$Criteria\ difficulty = \sum_{i=0}^{n} C_i \tag{1}$$

Where C is the cognitive thinking level assigned for each bloom keyword for a particular criterion.
n is the level of bloom's taxonomy

Figure 4 specifies the procedural steps for estimating the course difficulty.

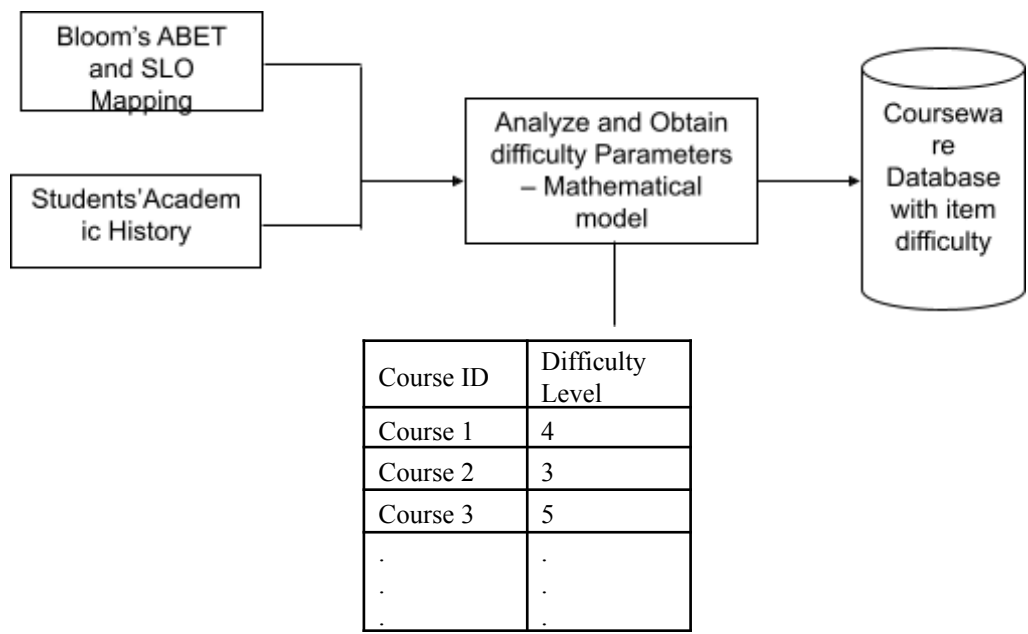

Figure 4: Course Difficulty Estimation

**Algorithm 1: Course Difficulty Estimation using ABET Student Outcomes and Blooms' Taxonomy**

Input:  B.Tech. CSE Curriculum with difficulty level as computed using Algorithm 2 and Algorithm 3
Output: Curriculum Courses with a Difficulty level
1.      Get the difficulty level of a course found from ABET Blooms' mapping and Student's Academic History using Algorithm 2.
2.      Identify the Mean Square Error for each difficulty parameter.

3.      Calculate the final difficulty level of a course

**Algorithm 2: Difficulty Estimation with Mapping Blooms Rubrics with ABET Mapping**

Input: Bloom's taxonomy and ABET Criteria
Output: Curriculum Courses with Difficulty level – Part I
1.      Set rubrics for Bloom's Taxonomy hierarchy as 1-6
2.      Map Blooms rubrics to ABET Student's Outcomes
3.      Mark every mapping of ABET with Bloom's Hierarchy value
4.      Repeat it for all ABET criteria
5.      Estimate the difficulty rubric for each ABET criteria by summing the Bloom's rubric mapped for each criterion.
6.      Level of Higher-order thinking is determined by the difficulty rubric. More the rubric value, more the difficulty level of the criteria is.

**Algorithm 3:** Difficulty Estimation Students' Academic History

Input: Student's academic history
Output: Courses with Difficulty level – Part II
1.      Get student details
2.      Get the courses registered by each student with grades for 3 generations
3.      Compare the mean for each generation.
4.      Average of three generations grades are considered as the difficulty parameter.

**Algorithm 4: Validating the Difficulty Estimation**

Input: Course Difficulty and student grades
Output: Comparison results
1.      Analyze the student's grades for each course
2.      If the course difficulty is high and the grades are low for more % of students, we infer that the difficulty of a course has an impact on the students' grades.

In the proposed method, the action words of bloom's are mapped with the ABET criteria, and for each criterion from a-m, the difficulty rubric is calculated using algorithm 2 Based on the grades of the students, the difficulty of a course is estimated using algorithm 3. Algorithm 1, compares the difficulty estimated from algorithm 2 and 3 and set a final difficulty rubric for each course.

## 4      Results and Discussions

The outcome mapped with each unit of a syllabus is mapped with the bloom's action words and are set with the rubrics as specified in Table 1. From the curriculum, 11 courses mapped with ABET criteria were chosen and applied with the rubric estimated as per Table 1. As per the specifications, the difficulty index of each course is estimated by identifying the number of ABET criteria mapped for each course with the bloom's total score 21 as specified in Table 2. This is part I estimation of difficulty index as per algorithm 2. The difficulty index is specified in the Likert scale from 1-5 from low to high.

**Table 2: Course Difficulty Estimation wrt. ABET Bloom mapping**

| Course Code /SO | a | b | c | d | e | f | g | h | i | j | k | l | m | Total ABET + Bloom Mapping | No. of criteria | Count *21 | Difficulty Index |
|---|---|---|---|---|---|---|---|---|---|---|---|---|---|---|---|---|---|
| Rubric | 6 | 21 | 21 | 6 | 21 | 3 | 3 | 6 | 21 | 1 | 6 | 21 | 21 | 157 | | | |
| C1 | 6 | 21 | | | 21 | | | | 21 | | 6 | 21 | | 96 | 6 | 126 | 3.8 |
| C2 | 6 | 21 | | | 21 | | | | 21 | | 6 | 21 | | 96 | 6 | 126 | 3.8 |
| C3 | | 21 | 21 | | | | | | 21 | | 6 | 21 | 21 | 111 | 6 | 126 | 4.4 |
| C4 | 6 | 21 | 21 | | 21 | | | | 21 | | | 21 | | 111 | 6 | 126 | 4.4 |
| C5 | 6 | 21 | | | 21 | | | | 21 | | 6 | 21 | 21 | 96 | 6 | 126 | 3.8 |
| C6 | 6 | 21 | | | 21 | | | | 21 | | 6 | 21 | | 96 | 6 | 126 | 3.8 |
| C7 | 6 | 21 | | | 21 | | | | 21 | | 6 | 21 | | 96 | 6 | 126 | 3.8 |
| C8 | | | 21 | | 21 | | | | | 6 | 6 | 21 | | 75 | 5 | 105 | 3.6 |
| C9 | 6 | | | | | | 5 | | | | 6 | 21 | | 38 | 4 | 84 | 2.3 |
| C10 | | 21 | | | 21 | | | 5 | 21 | | 6 | 21 | | 95 | 6 | 126 | 3.8 |
| C11 | | | | 6 | | 3 | 3 | | 6 | | | | | 18 | 4 | 84 | 1.1 |

The next part of difficulty estimation is by comparing the student's average grades scored in each course as specified in algorithm 2. The grades scored by 3 batches of students were considered for identifying the difficulty of a course.

$$DI = 5 - \left( \frac{Class\ Average}{100} \right) * 5 \qquad (2)$$

The class average is converted into 5 and subtracted with 5 for estimating the Difficulty Index (DI) as specified in equation (2). If the class average of a course is 35 out of 100, the difficulty index will be $5 - 1.75$, that is 3.25. Similarly, all the courses are estimated with be estimated for the difficulty index to the point of 5 as specified in Table 3. The actual DI and the predicted DI is depicted in figure 5.

Later, the estimated difficulty index with respect to bloom's ABET mappings are compared with the estimated difficulty index with respect to the student's grades and the mean square error is specified in Table 3.

**Table 3: Course Difficulty Estimation wrt. Student's Grades**

| Course Code /SO | Generation 1 | Generation 2 | Generation 3 | Student's Grades (Actual Difficulty) | ABET Bloom (Estimated Difficulty) | MSE |
|---|---|---|---|---|---|---|
| C1 | 4.2 | 3.4 | 4.4 | 4.0 | 3.8 | 0.2 |
| C2 | 4.2 | 4.1 | 3.8 | 4.0 | 3.8 | 0.2 |
| C3 | 4.3 | 3.9 | 4.2 | 4.1 | 4.4 | 0.3 |
| C4 | 3.9 | 4.2 | 4.5 | 4.2 | 4.4 | 0.2 |
| C5 | 3.9 | 3.8 | 4.3 | 4.0 | 3.8 | 0.2 |
| C6 | 4.1 | 4.3 | 3.9 | 4.1 | 3.8 | 0.3 |
| C7 | 3.6 | 3.4 | 3.8 | 3.6 | 3.8 | 0.2 |
| C8 | 3.4 | 3.8 | 3.6 | 3.6 | 3.6 | 0.0 |
| C9 | 2.4 | 2.6 | 2.1 | 2.4 | 2.3 | 0.1 |
| C10 | 4.2 | 3.9 | 4.1 | 4.1 | 3.8 | 0.3 |
| C11 | 1.6 | 1.2 | 1.4 | 1.4 | 1.1 | 0.3 |
| Average | | | | 3.6 | 3.5 | 0.2 |

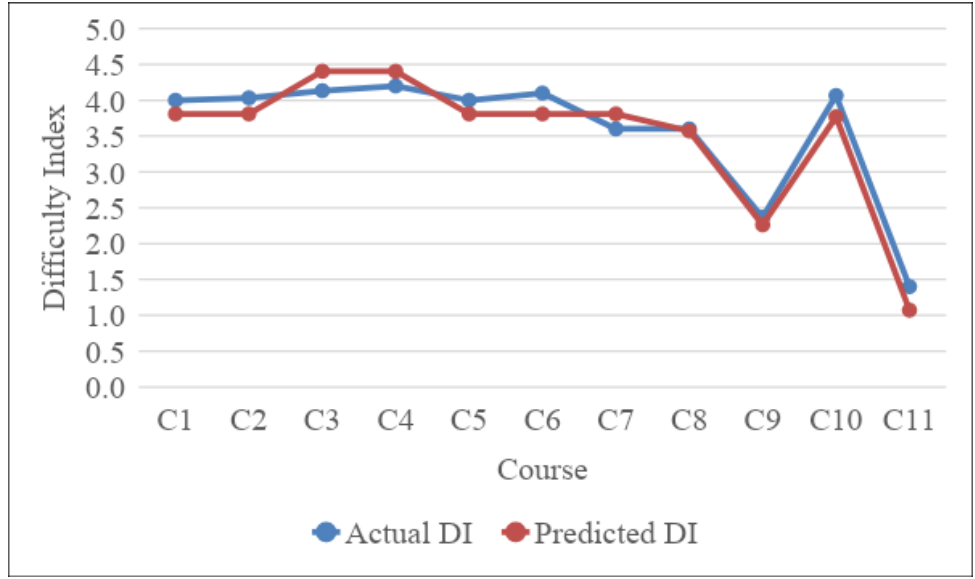

Figure 5: Actual DI Vs Predicted DI

## 5    Results and Discussions

The level and kind of training to be given to a course vary from course to course. The difficulty level of the course serves as one major parameter for deciding the number and type of tutoring to be done for the course. Hence, understanding and computing the difficulty level of the course may serve as a boom to student community which in-turn will directly impact the performance of the student in the examination. Bloom's taxonomy and ABET criteria set a mark in deciding quality of the system and hence difficulty estimation based on these two will obviously serve as a support for student and faculty. The estimated difficulty level of the courses is validated using grades of the appropriate courses for students belonging to three different batches. The accuracy which is the ratio of correct mapping of difficulty level to total courses is estimated to be 98%.